\documentclass{aastex7}

\begin{document}

\title{Magnetic Fields of Satellite Galaxies Stronger Than Comparable Centrals in TNG100}

\author[0000-0001-6928-4345]{Bryanne McDonough}
\email[show]{br.mcdonough@northeastern.edu}
\affiliation{Department of Physics, Northeastern University, 360 Huntington Ave, Boston, MA, USA}

\author[0009-0008-3349-0954]{Alexander Poulin}
\email{poulin.al@northeastern.edu}
\affiliation{Department of Physics, Northeastern University, 360 Huntington Ave, Boston, MA, USA}



\begin{abstract}

Magnetic fields exist in and around galaxies, but the properties of these fields have not been fully explored due to the challenges inherent in observing and modeling them. In this Note, we explore the differences in magnetic field strength of central and satellite galaxies from the magnetohydrodynamic TNG100 simulation. We find that on average, magnetic fields in satellite galaxies are roughly an order of magnitude stronger than those of central galaxies with comparable masses. The difference is greater for satellites that have already approached within $1 R_{200}$ of their host galaxies. These results indicate that magnetic fields in satellite galaxies are amplified by environmental processes as they fall into a host halo. 

\end{abstract}



\section{Introduction}
Theoretical results over the last decade have demonstrated that magnetic fields are an important component of the circumgalactic medium \citep[CGM; e.g.,][]{CGMBfld2023ARA&A..61..131F} and the intracluster medium \citep[ICM; e.g.,][]{ICMBFld2002ARA&A..40..319C}. However, the role magnetic fields play in galaxy evolution is poorly understood, in part due to inherent challenges in making observational measurements and the additional computational resources necessary to model magnetohydrodynamics (MHD) in simulations.

Magnetic fields may be particularly important for satellite galaxy evolution because they have been shown to increase the survival time of cool gas clouds in the circumgalactic medium \cite[e.g.,][]{McCourt2015MNRAS.449....2M} and magnetism in the ICM can support the survival of cool, dense cores of satellite halos that were subject to ram pressure stripping, resulting in “cold fronts” in X-ray observations \citep[e.g.,][]{markevitch2000ApJ...541..542M}. 
However, there have been few studies that have investigated magnetic fields in satellite galaxies. 

Recently, \cite{Werhahn2024arXiv240917229W} analyzed the magnetic fields of satellite galaxies in MHD zoom simulations of disk galaxies. They found that, compared to isolated dwarf galaxies with similar masses and star formation rates, satellite galaxies had stronger magnetic fields. While the precise mechanisms that act to amplify the satellite magnetic fields analyzed by \cite{Werhahn2024arXiv240917229W} remain unclear, they attribute this phenomenon to interactions with the CGM of the satellite's host because the amplification appears to occur during first infall.

In this Note, we demonstrate that the phenomenon of amplified magnetic fields is also found in satellite galaxies drawn from the full-volume TNG100 simulation \citep[e.g., ][]{TNG2017MNRAS.465.3291W,TNG2018MNRAS.480.5113M}. In Section \ref{sec:data}, we provide an overview of the TNG100 simulation and the data we obtain from it. In Section \ref{sec:results}, we compare the magnetic field strengths of simulated central and satellite galaxies and discuss how our results compare to those of \cite{Werhahn2024arXiv240917229W}. 

\section{Data} \label{sec:data}
Our sample of galaxies is drawn from the $z=0$ snapshot of the TNG100-1 (hereafter, TNG100) simulation. TNG100 is the $\sim100^3 \, {\rm Mpc}^3$ volume from the IllustrisTNG suite of cosmological MHD simulations. 
We restrict our analysis to subhalos (i.e., galaxies) that are cosmological in origin and have stellar masses $>10^9 M_\odot$. Here, we define stellar mass as the total mass of all stellar particles bound to the subhalo. Additionally, we restrict our analysis to include only galaxies with non-zero star formation rates (SFR) at $z=0$. We include this SFR restriction for consistency with the analysis of \cite{Werhahn2024arXiv240917229W}, although we do not find it significantly affects our results.

For each subhalo, the simulation reports a magnetic field strength that is computed using all gas cells bound to the subhalo by taking the square root of the volume-weighted sum of cells' squared magnetic field strengths, which gives a magnetic field strength with magnetic energy equivalent to the mean magnetic energy of all cells in the subhalo.

Within any friends-of-friends group (i.e., halo), the most massive subhalo is defined to be a central galaxy, while any less massive subhalos are satellites. We find $11,460$ central galaxies with anywhere between $0$ and $284$ satellites. Isolated centrals are those with no star-forming satellites, while non-isolated centrals have at least one child subhalo with non-zero SFR at $z=0$. In this Note, post-infall satellites are those that have approached the central galaxy of their host halo within $1R_{200}$ (i.e., the virial radius). Otherwise, satellites that have remained outside this radius are classified as pre-infall.

These selections result in a final count of $3648$ pre-infall, $202$ post-infall satellites, $9778$ isolated, and $1682$ non-isolated centrals.

\section{Results and Discussion} \label{sec:results}
\begin{figure}[ht!]
    \centering
    \includegraphics[width=0.95\linewidth]{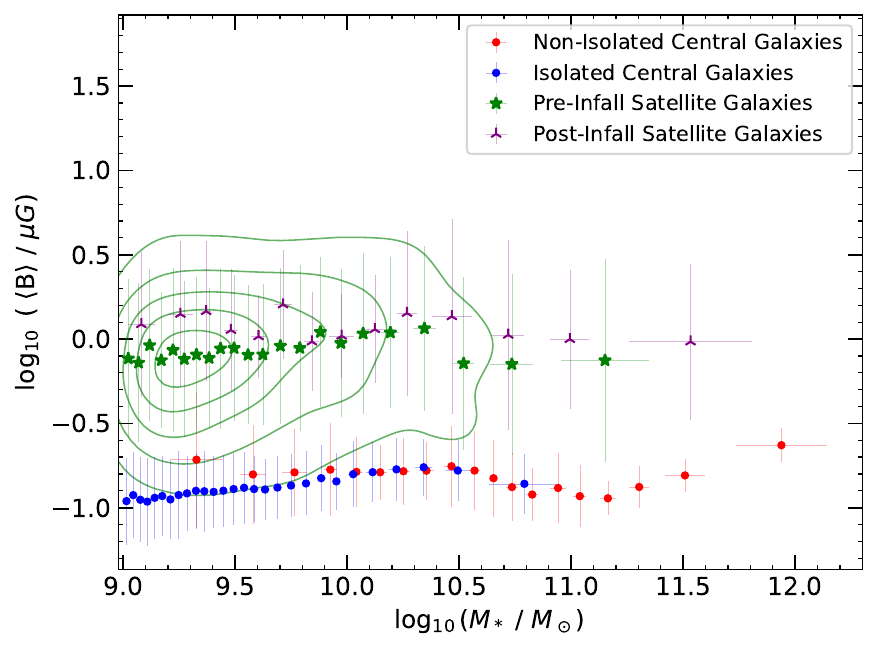}
    \caption{Magnetic field strength of galaxies as a function of stellar mass. Galaxies are binned by stellar mass, separated into pre-infall satellites (green), post-infall satellites (purple), non-isolated centrals (red), and isolated centrals (blue). Error bars on each point are the standard deviation in each bin. The contour lines represents the distribution of unbinned pre-infall satellites.}
    \label{fig:enter-label}
\end{figure}


Figure 1 gives the median magnetic field as a function of stellar mass, shown separately for isolated centrals (blue), non-isolated centrals (red), pre-infall satellites (green) and post-infall satellites (purple). Galaxies are binned by stellar mass such that each point in a group represents roughly the same number of galaxies ($\sim30$ galaxies per bin). Points represent the median values and error bars give the standard deviation. Finally, contours are drawn around the full distribution of pre-infall satellites to represent the wide variation within this relatively large group. 

While isolated and non-isolated centrals have similar median magnetic field strengths ($0.12\pm0.13$ $ \rm \mu G$ and $0.14\pm 0.30$ $\rm \mu G$, respectively), the median strengths of satellite magnetic fields are stronger, with differences for pre- and post- infall satellites. Post-infall satellites have the strongest median magnetic field strength of $1.3\pm2.1$ $\rm \mu G$, followed by pre-infall satellites with a median magnetic field strength of $0.83\pm3.4$ $\rm \mu G$. %

Individual galaxies can have magnetic field strengths that deviate significantly from the median for their bin, but the standard deviations for the satellites are greater than those for the centrals. We note that we do not discriminate between satellites that joined their hosts as central galaxies and those that joined their $z=0$ hosts as satellites of another system (i.e., satellites that were `pre-processed'). Some of these systems have magnetic field strengths comparable to those of central galaxies of similar masses. These may be satellites that have only joined their hosts recently, and have yet to experience the processes which result in amplification of their magnetic fields.


Overall, our results are in qualitative agreement with those of \cite{Werhahn2024arXiv240917229W}, in that satellite galaxies have, on average, stronger magnetic fields than comparable centrals. This is despite significant differences between our works. Our sample of galaxies from TNG100 is much larger, but they were simulated at a gas mass resolution of $m_{\rm baryon}\sim 10^6 M_\odot$ while the galaxies in \cite{Werhahn2024arXiv240917229W} were simulated at varying gas mass resolutions less than $5\times10^4 M_\odot$. Both simulations used the \texttt{Arepo} moving-mesh code to solve the magnetohydrodynamic equations, but there are differences in the applied galaxy formation models.

Additionally, differences arise in the classification of pre- and post- infall satellites. \cite{Werhahn2024arXiv240917229W} defined post-infall satellites as those at least $300$ Myr past the first pericenter of their orbit. In contrast, we define satellites as post-infall once they have approached within $1R_{200}$ of their host. Our results are not strongly affected by how infall is defined, although the separation between pre- and post- infall satellites decreases if the boundary is increased (e.g., from $1R_{200}$ to $3R_{200}$). 
We interpret this to mean that amplification of the magnetic fields of satellite galaxies increases as a satellite approaches its host galaxy.

While our results and those of \cite{Werhahn2024arXiv240917229W} are in agreement that satellite galaxies have stronger magnetic fields than centrals, we note that \cite{Werhahn2024arXiv240917229W} find a much stronger dependence on stellar mass for the magnetic field strength of each group. More work will be necessary to determine whether this difference arises due to differences in the galaxy formation models or simulation resolution.

The phenomenon of satellite galaxies having stronger magnetic fields than centrals has now been identified in two studies using different simulations. This pattern of amplification may be driven by environmental effects that a satellite experiences during its infall, including ram pressure and gravitational interactions. Stronger magnetic fields in satellite galaxies have implications for measurements of cosmic rays and studies into environmental impacts on satellite galaxy evolution. Thus, observational follow-up of these results and further investigations with simulations into the processes behind magnetic field amplification are very desirable.

In summary, this study finds that the magnetic fields of TNG100 satellite galaxies are amplified compared to central galaxies with similar masses, with greater amplification for satellite galaxies that have approached their host galaxy within $1R_{200}$.

\section{Acknowledgments}
The authors would like to thank the TNG team for the simulation data. BM acknowledges support from Northeastern University's Future Faculty Postdoctoral Fellowship Program.

%







\bibliography{sample631}{}
\bibliographystyle{aasjournalv7}



\end{document}